\documentclass[aps,prl,twocolumn,showpacs,superscriptaddress,floatfix]{revtex4-1}
\usepackage{graphicx,amsmath,amssymb}
\usepackage[usenames]{color}
\usepackage[dvipsnames]{xcolor}
\usepackage[unicode=true,pdfusetitle,
 bookmarks=false,
 breaklinks=false,pdfborder={0 0 1},backref=false,colorlinks,urlcolor=Black, linkcolor=blue,citecolor=blue]
 {hyperref}
\begin{document}
\title{Universal quantum Fisher information and simultaneous occurrence of Landau-class and topological-class transitions in non-Hermitian Jaynes-Cummings models
}
\author{Zu-Jian Ying }
\email{yingzj@lzu.edu.cn}
\affiliation{School of Physical Science and Technology, Lanzhou University, Lanzhou 730000, China}
\affiliation{Key Laboratory for Quantum Theory and Applications of MoE and Lanzhou Center for Theoretical Physics,
%and Key Laboratory of Theoretical Physics of Gansu Province, Lanzhou University,
Lanzhou University, Lanzhou 730000, China}

\begin{abstract}
Light-matter interactions provide an ideal testground for interplay of critical phenomena, topological transitions, quantum metrology and non-Hermitian physics.  We consider two fundamental non-Hermitian Jaynes-Cummings models which possess real energy spectra in parity-time (PT) symmetry and anti-PT symmetry. We show that the quantum Fisher information is critical around the transitions at the exceptional points and exhibits a super universality with respect to different parameters, all energy levels, both models, symmetric phases and symmetry-broken phases. The transitions are found to be both symmetry-breaking Landau-class transitions (LCTs) and symmetry-protected topological-class of transitions (TCTs), thus realizing a simultaneous occurrence of critical LCTs and TCTs which are conventionally incompatible due to contrary symmetry requirements.
\end{abstract}
\pacs{ }
\maketitle

%\date{\today}

With the theoretical
efforts~\cite{Braak2011,Solano2011,Boite2020,Liu2021AQT} and experimental progresses~\cite{Ciuti2005EarlyUSC,Aji2009EarlyUSC,
Diaz2019RevModPhy,Kockum2019NRP,Wallraff2004,Gunter2009, Niemczyk2010,Peropadre2010,FornDiaz2017,Forn-Diaz2010,Scalari2012,Xiang2013,Yoshihara2017NatPhys,Kockum2017,Bayer2017DeepStrong,Qin-ExpLightMatter-2018,LiPengBo-Magnon-PRL-2024}
over the past two decades, light-matter interactions have become a frontier field where simulations of traditional states of matter,
explorations of exotic quantum states and developments of quantum technologies meet and inspire novel sparks of ideologies.  In particular,
light-matter interactions can manifest few-body quantum phase transitions (QPTs)~\cite{
Liu2021AQT,Ashhab2013,Ying2015,Hwang2015PRL,Ying2020-nonlinear-bias,Ying-2021-AQT,LiuM2017PRL,Hwang2016PRL,Irish2017, Ying-gapped-top,Ying-Stark-top,Ying-Spin-Winding,Ying-2018-arxiv,Ying-JC-winding,Ying-Topo-JC-nonHermitian,Ying-gC-by-QFI-2024,
Grimaudo2022q2QPT,Grimaudo2023-Entropy,Zhu2024PRL}
which can be applied for
critical quantum metrology~\cite{Garbe2020,Garbe2021-Metrology,Ilias2022-Metrology,Ying2022-Metrology,Montenegro2021-Metrology,Hotter2024-Metrology,Chu2021-Metrology}.

Indeed, the continuous enhancements of coupling have brought the contemporary era of ultra-strong coupling\cite{Ciuti2005EarlyUSC,Aji2009EarlyUSC,Diaz2019RevModPhy,Kockum2019NRP,Wallraff2004,Gunter2009,
Niemczyk2010,Peropadre2010,FornDiaz2017,Forn-Diaz2010,Scalari2012,Xiang2013,Yoshihara2017NatPhys,Kockum2017,
Ulstrong-JC-2,Ulstrong-JC-3-Adam-2019} and deep-strong coupling~\cite{Yoshihara2017NatPhys,Bayer2017DeepStrong,Ulstrong-JC-1}. An intriguing
phenomenon in the emerging phenomenology\cite{Braak2011,Solano2011,Boite2020,Liu2021AQT,
Ashhab2013,Ying2015,Hwang2015PRL,Ying2020-nonlinear-bias,Ying-2021-AQT,LiuM2017PRL,Hwang2016PRL,Irish2017, Ying-gapped-top,Ying-Stark-top,Ying-Spin-Winding,Ying-2018-arxiv,Ying-JC-winding,Ying-Topo-JC-nonHermitian,Ying-gC-by-QFI-2024,
Grimaudo2022q2QPT,Grimaudo2023-Entropy,Zhu2024PRL,
Ulstrong-JC-3-Adam-2019,Ulstrong-JC-2,Ulstrong-JC-1,Qin-ExpLightMatter-2018,LiPengBo-Magnon-PRL-2024,
PRX-Xie-Anistropy,GongMing-2021,Padilla2022,Wolf2012,FelicettiPRL2020,Simone2018,Alushi2023PRX,
 Garbe2020,Garbe2021-Metrology,Ilias2022-Metrology,Ying2022-Metrology,Montenegro2021-Metrology,Hotter2024-Metrology,
 Irish-class-quan-corresp,Eckle-2017JPA,Cong2022Peter,Zhu-PRL-2020,
 Felicetti2018-mixed-TPP-SPP,Felicetti2015-TwoPhotonProcess,e-collpase-Garbe-2017,Rico2020,
 Boite2016-Photon-Blockade,Ridolfo2012-Photon-Blockade,Li2020conical,
 Liu2015,CongLei2017,CongLei2019,LiuGang2023,Ma2020Nonlinear,ZhangYY2016,
 ChenQH2012,e-collpase-Duan-2016,ZhengHang2017,Zheng2017,Yan2023-AQT,
 Chen-2021-NC,Lu-2018-1,Gao2021,Batchelor2015,XieQ-2017JPA,Bera2014Polaron,ChenGang2012,FengMang2013,PengJie2019,Casanova2018npj,
 Braak2019Symmetry,HiddenSymMangazeev2021,HiddenSymLi2021,HiddenSymBustos2021,
 Stark-Cong2020,Stark-Grimsmo2013,Downing2014,Pietikainen2017,Yimin2018,Wang2019Anisotropy,you024532,Chen2018,WangZJ2020PRL,
 Bermudez2007} of ultra-strong and deep-strong couplings is the existence\cite%
{Liu2021AQT,Ashhab2013,Ying2015,Hwang2015PRL,Ying2020-nonlinear-bias,Ying-2021-AQT,LiuM2017PRL,Hwang2016PRL,Irish2017,
Ying-gapped-top,Ying-Stark-top,Ying-2018-arxiv,Ying-Spin-Winding,Ying-JC-winding,Ying-gC-by-QFI-2024,Grimaudo2022q2QPT}
of a QPT in the fundamental models of light-matter interactions, the quantum Rabi
model (QRM)\cite{rabi1936,Rabi-Braak,Eckle-Book-Models} and the
Jaynes-Cummings model (JCM)\cite{JC-model,JC-Larson2021} which are few-body systems, while traditionally QPTs
lie in condensed matter\cite{Sachdev-QPT}. Here in light-matter interactions, the QPT occurs
in the low-frequency limit which is a replacement of thermodynamical limit
in many-body systems. In such a limit, the QPT exhibits a scaling behavior and forms critical
universality, which is also a character often born with QPTs in the condensed matter\cite{Sachdev-QPT,Irish2017}. Such critical universality not is only valid
for anisotropy\cite{LiuM2017PRL,Ying-2021-AQT} but also holds for the Stark non-linear
coupling\cite{Ying-Stark-top}. Moreover, the critical exponents can be
bridged to the thermodynamical case\cite{LiuM2017PRL}. It is noticed that the QPT here has a hidden symmetry breaking
which characterizes the traditional Landau class of transition (LCT)\cite{Ying-2021-AQT,Ying-Stark-top,Landau1937}.

When the frequency is tuned up, a series of different phase transitions emerge apart from the QPT in the low-frequency limit~\cite{Ying-2021-AQT,Ying-gapped-top,Ying-Stark-top}.
In such a finite-frequency situation, the critical universality breaks down and the system properties are diversified. Surprisingly, a reformed universality is revealed in such diversified situation as each phase shares a common number of nodes in the wavefunction~\cite{Ying-2021-AQT,Ying-gapped-top,Ying-Stark-top} which are further found to correspond to spin winding~\cite{Ying-Spin-Winding,Ying-JC-winding,Ying-Topo-JC-nonHermitian}. These topological features endow the emerging transitions the connotation of topological-class of transition (TCT),
%including the conventional TCTs~\cite{Ying-2021-AQT} and unconventional ones~\cite{Ying-gapped-top,Ying-Spin-Winding,Ying-JC-winding,Ying-Topo-JC-nonHermitian},
analogously to the TCTs in condensed matter~\cite{
Topo-KT-transition,
Topo-KT-NoSymBreak,
Topo-Haldane-1,Topo-Haldane-2,
Topo-Wen,ColloqTopoWen2010,
Hasan2010-RMP-topo,Yu2010ScienceHall,Chen2019GapClosing,TopCriterion,Top-Song-2017,Top-Guan,TopNori,
Amaricci-2015-no-gap-closing,Xie-QAH-2021},
which preserves %or is protected by
the symmetry in contrast to the symmetry-breaking character in LCT. Also, the collapsed critical universality is replaced by topological universality~\cite{Ying-Stark-top}. The topological feature is also found to be robust against the non-Hermiticity~\cite{Ying-Topo-JC-nonHermitian} arising from dissipation and decay rates~\cite{NonHermitianJCM-2022SciChina}.

Note here that the TCTs and LCT occur at the same system of light-matter interaction, which raises the issue of coexistence of the TCT and LCT while they are conventionally incompatible due to the contrary symmetry requirements~\cite{Ying-2021-AQT,Ying-Stark-top,Ying-JC-winding}. The anisotropic QRM has both the TCTs and LCT  but at different couplings~\cite{Ying-2021-AQT,Ying-Stark-top}. The JCM has a coexistence of the TCT and LCT, however the LCT is of first order at finite frequencies thus not critical~\cite{Ying-JC-winding}. It is desirable to obtain a simultaneous occurrence of critical TCT and LCT, which would not only yield a conceptional upgrade for the TCT-LCT coexistence, but also have both advantages of sensitive critical feature and robust topological feature
at the same time.

In this work we consider two non-Hermitian JCMs which have real energy spectra in  parity-time (PT) and anti-PT symmetries, with spontaneous symmetry breaking transitions at exceptional points. We show that the quantum-metrology-related quantum Fisher information (QFI)~\cite{Garbe2020,Garbe2021-Metrology,Ilias2022-Metrology,Ying2022-Metrology,Montenegro2021-Metrology,Hotter2024-Metrology,AnJH-Metrology-noise-2023} is critical and universal.
%exhibits a super universality with respect to different parameters,  all energy levels, both models, symmetric phases and symmetry-broken phases.
The transitions are found to be also TCTs, thus realizing a simultaneous occurrence of critical TCTs and LCTs.

{\it Non-Hermitian JCMs.--}We start with the generic non-Hermitian JCM~\cite%
{Qin-ExpLightMatter-2018,NonHermitianJCM-2022SciChina,JC-model,JC-Larson2021,Ulstrong-JC-2}
\begin{equation}
H=\widetilde{\omega }a^{\dagger }a+\frac{\widetilde{\Omega }}{2}\sigma _{x}+%
\widetilde{g}\left( \widetilde{\sigma }_{-}a^{\dagger }+\widetilde{\sigma }%
_{+}a\right)   \label{H}
\end{equation}%
which describes the coupling between a quantized bosonic mode created
(annihilated) by $a^{\dagger }$ ($a)$ and a qubit denoted by the Pauli
matrices $\sigma _{x,y,z}$. Here $\sigma _{z}=\pm $ represents the
two flux states in the flux-qubit circuit system~\cite{Irish2014,flux-qubit-Mooij-1999}
and $\widetilde{\sigma }^{\pm }=(\sigma _{z}\mp
i\sigma _{y})/2$ raises and lowers the spin states $\Uparrow ,\Downarrow $
on the $\sigma _{x}$ basis. The bosonic frequency, qubit energy splitting
and coupling strength are complex\cite{NonHermitianJCM-2022SciChina}
$\widetilde{\omega }=\omega -i\kappa$, $\widetilde{\Omega }=\Omega
-i\gamma$, $\widetilde{g}=g-i\Gamma$, due to the dissipation and decay
rates\cite{NonHermitianJCM-2022SciChina}. The non-Hermitian Hamiltonian $H$
has the same eigenvectors as the Liouvillian in the Lindblad master equation%
\cite{NonHermitian-Nori2019,Plenio1998-quantum-jump} in the negligible
quantum jump term\cite{NonHermitianJCM-2022SciChina}. The eigenstate has the
form $\psi _{n}^{( \eta ) }=(
C_{n\Uparrow }^{( \eta )}\left\vert n-1,\Uparrow \right\rangle
+C_{n\Downarrow }^{(\eta)}\left\vert n,\Downarrow \right\rangle )
/
\sqrt{N_{n}^{(\eta)}}$,
where $C_{n\Uparrow }^{(\eta )}=e_{-}+\eta \sqrt{e_{-}^{2}+n\ \widetilde{g}^{2}}$,
$C_{n\Downarrow }^{\left( \eta \right) }=\widetilde{g}\sqrt{n},e_{+}=\left( n-\frac{1}{2}\right) \widetilde{\omega }$,
$e_{-}=\frac{1}{2}(\widetilde{\Omega }-\widetilde{\omega }) $
and $N_{n}^{( \eta ) }=|C_{n\Uparrow }^{(\eta)}|^{2}+|C_{n\Downarrow }^{(\eta ) }|^{2}$. Here $\eta =\pm $
labels two energy branches
$E_n^{\left(\eta \right) } = e_{+}+\eta \sqrt{e_{-}^{2}+n\ \widetilde{g}^{2}
}$
and $n$ denotes photon number, except
$\psi _{0}=\left\vert 0,\Downarrow \right\rangle $ and $E_{0}=-\frac{\widetilde{\Omega }}{2}$ for $n=0$.
%%%%%%%%%%%%%%%%%%%%%%%%%%%%%%%%%%%%%%%%%%%%%%%%%%%%%%%%%%%%%%%%%%%%%%%%%%%%%%%%%%%%%%%%%%%%%%%%%%
\begin{figure}[t]
\includegraphics[width=1\columnwidth]{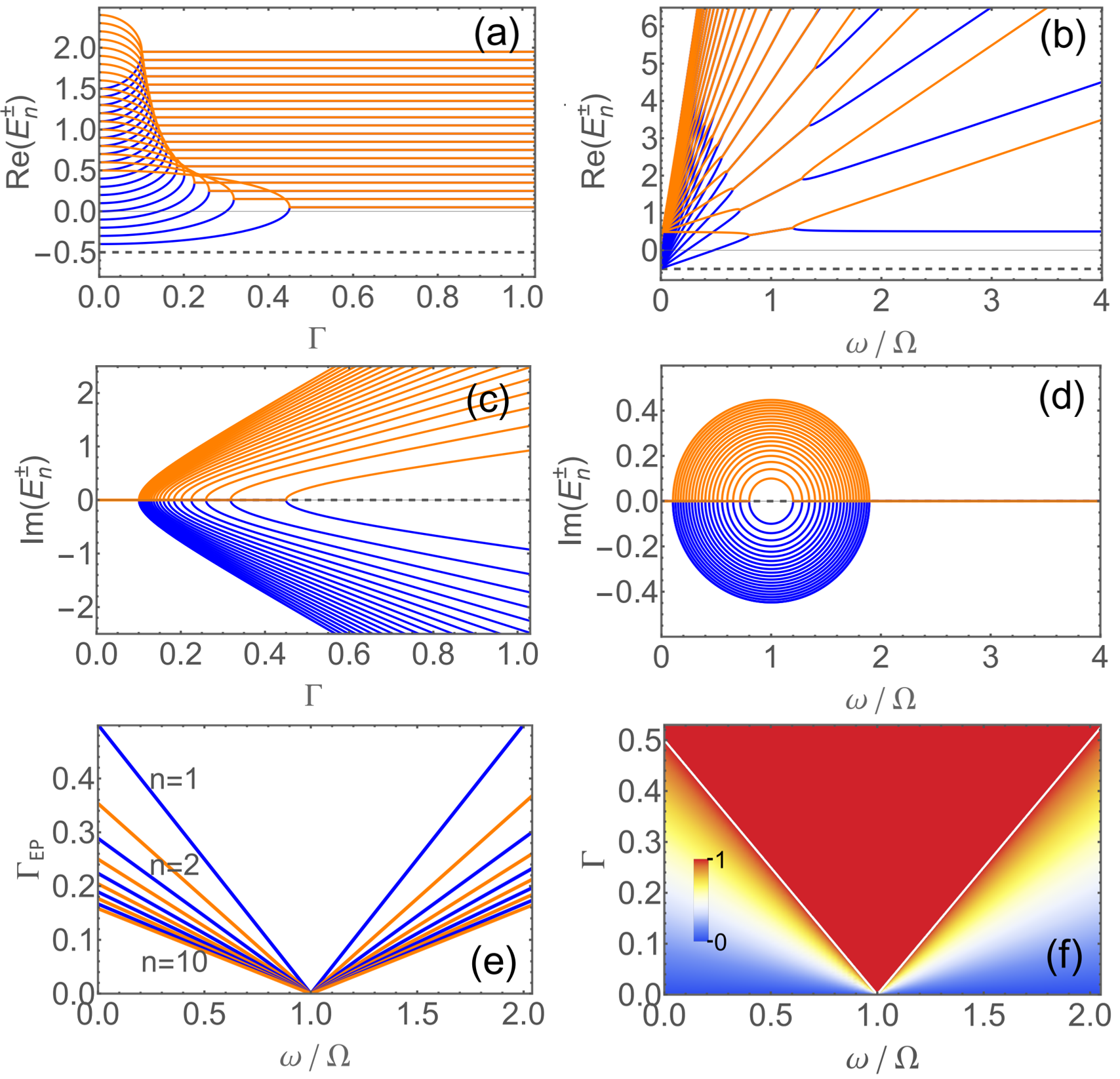}
\caption{Real part (a,b) and imaginary part (c,d) of the energy spectrum versus $\Gamma$ (a,c) and $\omega$ (b,d) for $H_\Gamma$. (e) Exceptional points (EPs) in different levels $n$. (f) Expectation amplitude $|\langle\Pi_xK\rangle|$ in the $\omega$-$\Gamma$ plane for $H_\Gamma$. $H_\gamma$ has similar splitting of the imaginary part of the energy spectrum, EPs and map of $\omega$-$\Gamma$ as in (c),(e) and (f) with $\Gamma$ and $|\omega-\Omega|$ replaced by $\gamma$ and $g$. Here, $\omega=0.1\Omega$ in (a) and (c) and $\Gamma=0.1\Omega$ in (b) and (d), $n=1$ in (f), and $\Omega=1$ is set as the units in all panels.
}
\label{fig-ReE-ImE}
\end{figure}
%%%%%%%%%%%%%%%%%%%%%%%%%%%%%%%%%%%%%%%%%%%%%%%%%%%%%%%%%%%%%%%%%%%%%%%%%%%%%%%%%%%%%%%%%%%%%%%%%%
In the following, we focus on two special cases%
\begin{eqnarray}
H_{\gamma } &=&\omega a^{\dagger }a+\frac{\Omega _{\gamma }-i\gamma }{2}%
\sigma _{x}+g\left( \widetilde{\sigma }_{-}a^{\dagger }+\widetilde{\sigma }%
_{+}a\right) ,  \label{H-gamma} \\
H_{\Gamma } &=&\omega a^{\dagger }a+\frac{\Omega }{2}\sigma _{x}+i\Gamma
\left( \widetilde{\sigma }_{-}a^{\dagger }+\widetilde{\sigma }_{+}a\right) ,
\end{eqnarray}%
with $\Omega _{\gamma }=\omega $, which have real energy spectra and exhibit
universal QFI and simultaneous occurrence of LCTs and
TCTs.

{\it Exceptional point (EP) and PT/anti-PT symmetry breaking.}--We introduce
a unitary operator
\begin{eqnarray}
\Pi _{x} &=&[ a( a^{\dagger }a ) ^{-1/2}\widetilde{\sigma }
_{+}+( a^{\dagger }a) ^{-1/2}a^{\dagger }\widetilde{\sigma }_{-}
],
\end{eqnarray}
which exchanges the basis ${\cal B}_{n}=\left\vert n-1,\Uparrow
\right\rangle $ and $\left\vert n,\Downarrow \right\rangle $, and the
conjugate operator $K$: $i\rightarrow -i,$ whose product $\Pi _{x}K$ forms a
conventional representation of PT symmetry in non-Hermitian physics~\cite{NonHermitian-Bender2007,NonHermitian-Bergholtz2021,NonHermitian-Review-Ashida2020,
EP-sensors-Li-2023,EP-sensors-Miri-2019,EP-sensors-Ozdemir-2019,EP-sensors-Hokmabadi-2019,EP-sensors-Chen-2017,EP-sensors-Wiersig-2014,EP-sensors-Liu-2016,
Non-Hermitian-JinLiang-2019,Non-Hermitian-ChenShu-2021-PRL,Non-Hermitian-Yao-2018,Non-Hermitian-Yokomizo-2019,Non-Hermitian-Gong-2018}. One
can also adopt the time reversal operator with spin, $T=i\sigma _{y}K$, and
define another parity operator $P=\Pi _{x}\left( -i\sigma _{y}\right) $. The
model $H_{\gamma }$ has the PT symmetry%
\begin{equation}
\Pi _{x}KH_{\gamma }K^{-1}\Pi _{x}^{-1}=H_{\gamma }
\end{equation}%
while $H_{\Gamma }$ has an anti-PT symmetry
\begin{equation}
K\Pi _{x}H_{\Gamma }\Pi _{x}^{-1}K^{-1}=-H_{\Gamma }+\omega (2a^{\dagger }a+
\sigma _{x}) \label{anti-PT}
\end{equation}%
as the second term becomes a constant in the ${\cal B}_{n}$ subspace due to the U(1) symmetry $a^{\dagger }a+\sigma _{x}/2+1/2=n$. Although
conventional anti-PT symmetry does not bring a constant, this extended
anti-PT symmetry \eqref{anti-PT} can also lead to real energy spectrum. Indeed,
the imaginary part of the energy, ${\rm Im}(E_{n}^{(\pm )})$, vanishes at the EPs
\begin{eqnarray}
g_{{\rm EP}} &=&\gamma /(2\sqrt{n}),\quad \gamma _{{\rm EP}}=2\sqrt{n}g; \\
\omega _{{\rm EP}\pm } &=&\Omega \pm 2\sqrt{n}\Gamma ,\quad \Gamma _{{\rm EP}%
\pm }=\pm (\Omega -\omega )/(2\sqrt{n}),
\end{eqnarray}%
where the eigenenergies and eigenstates coalesce. We
illustrate the case for $H_{\Gamma }$ in Fig. \ref{fig-ReE-ImE}, with the
real and imaginary parts of $E_{n}^{(\pm )}$ in panels (a)-(d) and the
EPs in (e). Although the Hamiltonians $H_{\gamma }$ and $H_{\Gamma }$
possess the PT and anti-PT symmetries respectively, the eigenstates have
spontaneous symmetry breaking at the EPs, as indicated by the expectation
amplitude $\left\vert \langle K\Pi _{x}\rangle \right\vert $ in Fig. \ref%
{fig-ReE-ImE}(f). Here the eigenstate is anti-PT-symmtric in the red region
above the EP (white lines) but becomes anti-PT-broken in the
region below the EP, with $\langle K\Pi _{x}\rangle =1$ in the former and $%
\left\vert \langle K\Pi _{x}\rangle \right\vert =2\sqrt{n}\Gamma /\left\vert
\Omega -\omega \right\vert < 1$ in the latter. Real
energy spectrum also appears in $H_{\gamma }$ similarly by replacing \{$\Gamma ,\omega -\Omega $\} with \{$%
\gamma ,g$\}. However, $H_{\gamma }$ has real (complex) energy spectrum in
the PT-symmetric (PT-broken) phase while $H_{\Gamma }$ does reversely in the
anti-PT-broken (anti-PT-symmetric) phase, as later compared in Table \ref{Table-LCT-TCT}.

%%%%%%%%%%%%%%%%%%%%%%%%%%%%%%%%%%%%%%%%%%%%%%%%%%%%%%%%%%%%%%%%%%%%%%%%%%%%%%%%%%%%%%%%%%%%%%%%%%
\begin{figure}[t]
\includegraphics[width=1\columnwidth]{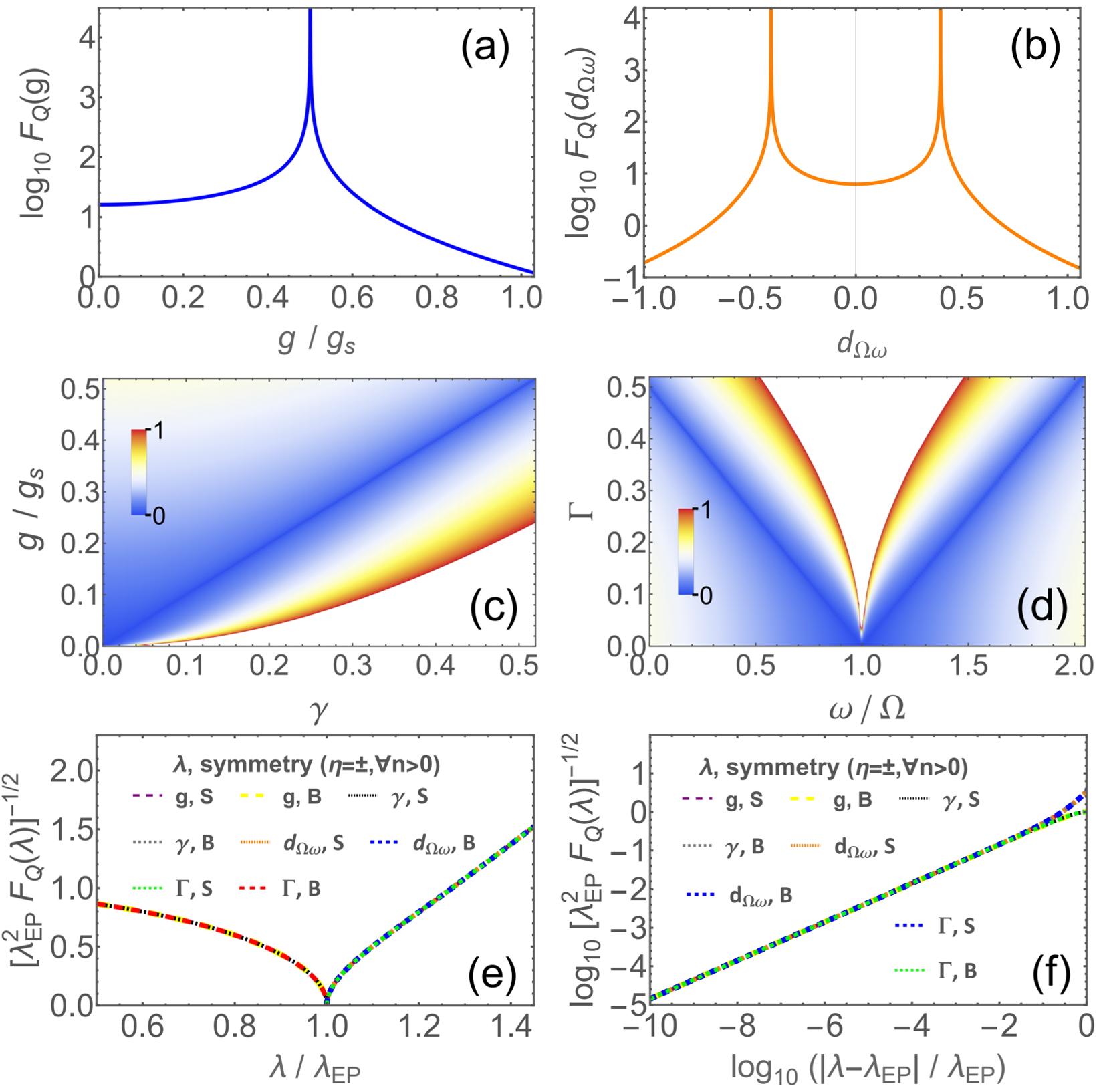}
\caption{Critical and universal quantum Fisher information (QFI) around the EPs.
(a) $log_{10}F_Q(g)$ versus $g$ at $\gamma =0.5\Omega$ in $H_\gamma$.
(b) $log_{10}F_Q(d_{\Omega\omega})$ versus $d_{\Omega\omega}=\Omega-\omega$ at $\omega=0.1\Omega$ in in $H_\Gamma$.
(c) $F_Q(\gamma)^{-1/2}$ in the $\gamma$-$g$ plane for $H_\gamma$.
(d) $F_Q(\gamma)^{-1/2}$ in the $\omega$-$\Gamma$ plane for $H_\Gamma$.
(e) Scaling relation of $[\gamma_{\rm EP}^2F_Q(\lambda)]^{-1/2}$ versus $\lambda/\lambda _{\rm EP}$ for all $\lambda =\gamma ,g,\Gamma$ and $\omega $ and all energy levels \{$\eta,n$\} of $H_\gamma$ and $H_\Gamma$ in symmetric (B) phases and symmetry-broken (B) phases.
(f) Scaling relation of $[\gamma_{\rm EP}^2F_Q(\lambda)]^{-1/2}$ versus $|\lambda-\lambda _{\rm EP}|/\lambda _{\rm EP}$ in logarithm scale around the EPs for all $\lambda =\gamma ,g,\Gamma$ and $\omega $ and all energy levels \{$\eta,n$\}. In (a) $g$ is scaled by $g_{\rm s}=\sqrt{\omega \Omega}/2$, while $\Omega=1$ is set as the units of other parameters.
}
\label{fig-QFI}
\end{figure}
%%%%%%%%%%%%%%%%%%%%%%%%%%%%%%%%%%%%%%%%%%%%%%%%%%%%%%%%%%%%%%%%%%%%%%%%%%%%%%%%%%%%%%%%%%%%%%%%%%

{\it Critical and universal QFI.} The critical behavior of the QPT in light-matter interactions has been applied for the critical
quantum metrology \cite%
{Garbe2020,Garbe2021-Metrology,Ilias2022-Metrology,Ying2022-Metrology}. Although the
QPT occurs in Hermitian case, here in the
non-Hermitian JCMs we also see critical behavior around the EPs. In
quantum metrology the precision of experimental estimation of a parameter $\lambda $
in the Hamiltonian is bounded by $F_{Q}^{1/2}$\cite%
{Cramer-Rao-bound}, where $F_{Q}$ is the QFI
\cite{Cramer-Rao-bound,Taddei2013FisherInfo,RamsPRX2018} which takes the
following form for a pure state $\psi(\lambda)$
\begin{equation}
F_{Q}\left( \lambda \right) =4[ \langle \psi ^{\prime }\left( \lambda
\right) |\psi ^{\prime }\left( \lambda \right) \rangle -\left\vert \langle
\psi ^{\prime }\left( \lambda \right) |\psi \left( \lambda \right) \rangle
\right\vert ^{2}] ,
\end{equation}%
where $^{\prime }$ denotes the derivative with respect to $\lambda $. A higher QFI means a higher
measurement precision. $F_{Q}$ is equivalent to the susceptibility of the
fidelity whose critical behavior characterizes
QPT\cite{Zhou-FidelityQPT-2008,Zanardi-FidelityQPT-2006,Gu-FidelityQPT-2010,You-FidelityQPT-2007,You-FidelityQPT-2015}.

Here we see that the QFI exhibits the critical character and obeys a
universal relation. Indeed, as illustrated by Figs.\ref{fig-QFI}(a) and \ref{fig-QFI}(b),
the QFI with respect to the parameters $\gamma ,g,\Gamma$, or $\omega$ is diverging around the EPs.
The diverging behavior is also
reflected by the vanishing inverse QFI $F_{Q}^{-1/2}$ which is the ultimate
bound of experimental measurement errors, as shown by the 2D maps in Figs.\ref{fig-QFI}(c) and \ref{fig-QFI}(d)
where $F_{Q}^{-1/2}$ becomes zero at the EPs
(blue diagonal lines). We find that the QFI with respect to
different parameters follows a same scaling relation, as demonstrated by the
Fig.\ref{fig-QFI}(e) where the behaviors of $F_{Q}^{-1/2}$ for $\gamma
,g,\Gamma $ and $\omega $ with different levels number $n$ and $\eta $
collapse onto a same line on both sides of the EP. Moreover, both sides have
the same critical exponent around the
EPs, as displayed by Fig.\ref{fig-QFI}(f). As a matter of fact, the QFI takes the form
\begin{equation}
F_{Q}\left( \lambda \right) =\frac{\theta \left( \lambda _{{\rm EP}}-\lambda
\right) }{\lambda _{{\rm EP}}^{2}-\lambda ^{2}}+\frac{\lambda _{{\rm EP}%
}^{2}\theta \left( \lambda -\lambda _{{\rm EP}}\right) }{\lambda ^{2}\left(
\lambda ^{2}-\lambda _{{\rm EP}}^{2}\right) },  \label{Fq-lambda-1}
\end{equation}%
universally for any $\lambda =\gamma ,g,\Gamma $ or $\omega $, despite the
different roles they play in the system. Here $\theta \left( x\right) $ is
the Heaviside step function. Around the EPs on both sides of the EPs the QFI
can be unified as%
\begin{equation}
F_{Q}\left( \lambda \right) =\frac{\left\vert \lambda -\lambda _{{\rm EP}%
}\right\vert ^{-1}}{2\lambda _{{\rm EP}}}.  \label{Fq-lambda-2}
\end{equation}%
Note that Eqs. (\ref{Fq-lambda-1}) and (\ref{Fq-lambda-2}) are independent
of the energy level number $n$ and energy branch marker $\eta $, thus being
universal also for all levels.

%%%%%%%%%%%%%%%%%%%%%%%%%%%%%%%%%%%%%%%%%%%%%%%%%%%%%%%%%%%%%%%%%%%%%%%%%%%%%%%%%%%%%%%%%%%%%%%%%%
\begin{figure}[t]
\includegraphics[width=1\columnwidth]{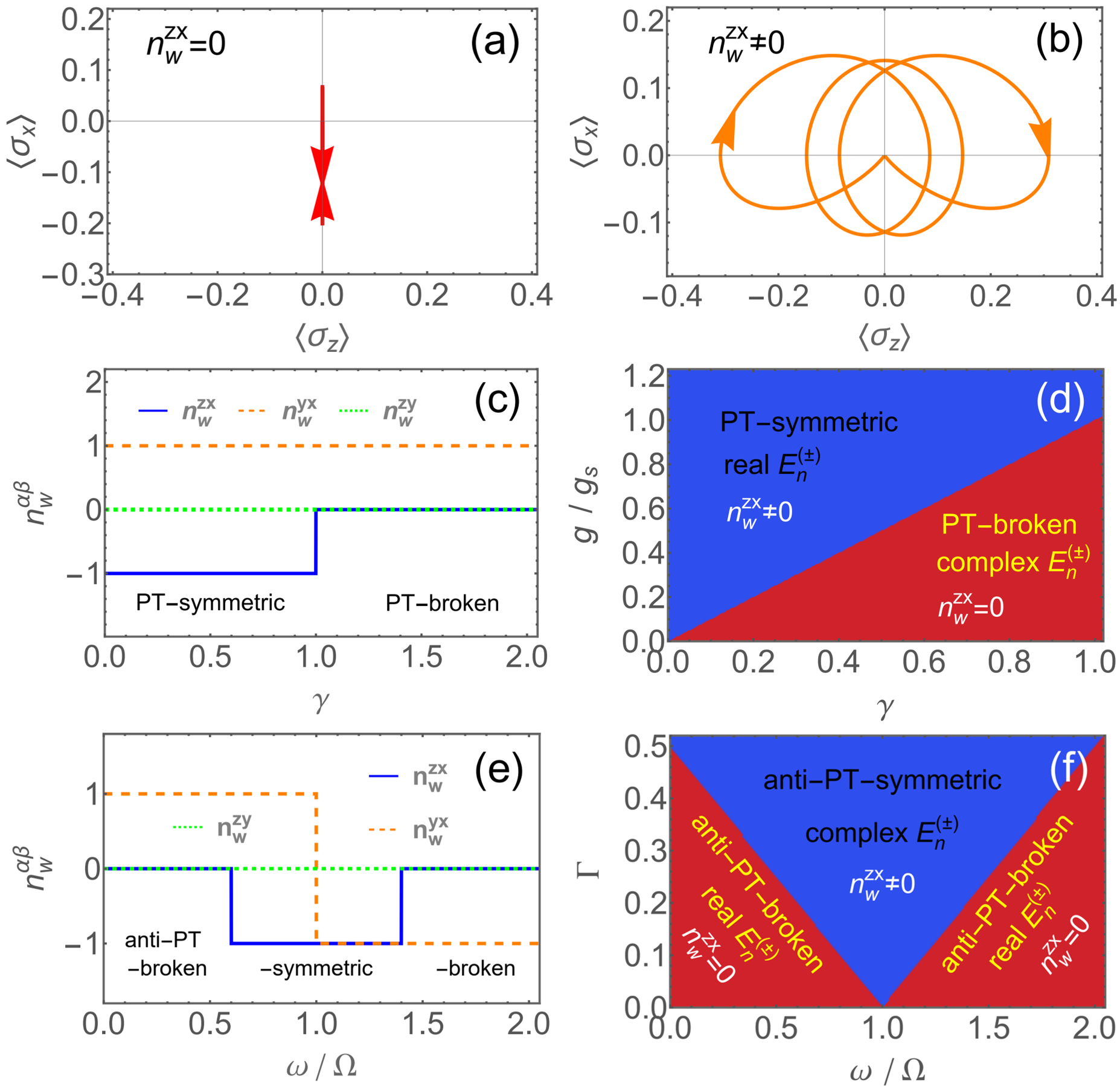}
\caption{Topological transitions at the EPs.  (a) Zero spin winding in the $\langle \sigma _z \rangle$-$\langle \sigma _x \rangle$ plane in PT-/anti-PT-broken phase phases. (b) Finite spin winding in PT-/anti-PT-symmetric phases. (c) Spin winding number $n_w^{\alpha\beta}$ in different planes versus $\gamma$ at $g=1.0g_{\rm s}$ for $H_\gamma$. (d) Spin winding number $n_w^{zx}$ in the $\gamma$-$g$ plane with PT/anti-PT and $E_n^{\pm}$ situations. (e) $n_w^{\alpha\beta}$ versus $\omega$ at $\Gamma=0.2\Omega$ for $H_\Gamma$. (f) $n_w^{zx}$ in the $\omega$-$\Gamma$ plane. Here, as an illustration, $n=3$ in (a) and (b) and $n=1$ in (c)-(f).
}
\label{fig-topo-transition}
\end{figure}
%%%%%%%%%%%%%%%%%%%%%%%%%%%%%%%%%%%%%%%%%%%%%%%%%%%%%%%%%%%%%%%%%%%%%%%%%%%%%%%%%%%%%%%%%%%%%%%%%%

\begin{table}[ht]
\caption{Comparison of critical and topological features in the PT-symmetric (PT-S) and PT-broken (PT-B) phases of $H_\gamma$ and the anti-PT-symmetric (anti-PT-S) and anti-PT-broken (anti-PT-B) phases of $H_\Gamma$. }
\centering % used for centering table
\begin{tabular}{c|c|c||c|c}
\hline\hline
Parameter    &\multicolumn{2}{|c||} {$H_\gamma$}             &\multicolumn{2}{|c} {$H_\Gamma$}   \\[0.5ex]
             \cline{1-5}
phase        &PT-S            &PT-B        &anti-PT-S      & anti-PT-B \\[0.5ex]
\hline
$E_n^{(\pm)}$        &real            & complex     & complex        &real \\
Re($E_n^{(\pm)}$)    & splitting      & degenerate  & degenerate     & splitting \\
Im($E_n^{(\pm)}$)    & degenerate     & splitting   & splitting      & degenerate \\
\hline
transition   &\multicolumn{2}{|c||} {Landau-class} &\multicolumn{2}{|c} {Landau-class}\\
             \cline{1-5}
$\Pi _{x}K$  & preserved      & broken      & preserved      & broken \\
QFI          & universal      & universal   & universal      & universal \\
criticality  & critical       & critical    & critical       & critical \\
%order para.     & varying        & varying     & varying        & varying \\
\hline
transition   &\multicolumn{2}{|c||} {topological-class} &\multicolumn{2}{|c} {topological-class}\\
             \cline{1-5}
$e^{i\pi a^{\dagger }a}\sigma _{x}$  & preserved   & preserved      &  preserved           &  preserved \\
$n_w^{zx}$   & nonzero        & zero        & zero           & nonzero  \\
topo. no.      & invariant      & invariant   & invariant      & invariant \\
[1ex] \hline
\end{tabular}
\label{Table-LCT-TCT}% is used to refer this table in the text
\end{table}

{\it Topological transitions at the EPs}.--Representing the photon number
state $\left\vert n\right\rangle $ by the eigenfunction of harmonic
orscillator $\phi _{n}\left( x\right) =\frac{1}{\pi ^{1/4}\sqrt{2^{n}n!}}%
H_{n}\left( x\right) e^{-x^{2}/2}$ with the Hermite polynomial $H_{n}\left(
x\right) $, the eignfunction on the $\sigma _{z}=\uparrow
,\downarrow $ basis becomes $\psi _{n}^{\left( z,\eta \right) }=\ \psi
_{+}^{z}\left( x\right) \left\vert \uparrow \right\rangle +\psi
_{-}^{z}\left( x\right) \left\vert \downarrow \right\rangle $ where $\psi
_{\pm }^{z}(x) =[C_{n\Uparrow }^{\left( \eta \right) }\phi _{n-1}\left(
x\right) \pm C_{n\Downarrow }^{\left( \eta \right) }\phi _{n}\left( x\right)
]/\sqrt{2N_{n}^{(\eta )}}$. The spin textures are determined by
$
\langle \sigma _{z}\left( x\right) \rangle  = \left\vert \psi
_{+}^{z}\left( x\right) \right\vert ^{2}-\left\vert \psi _{-}^{z}\left(
x\right) \right\vert ^{2}  \label{SpinZ-byWave}
$,
$
\langle \sigma _{x}\left( x\right) \rangle  =\left\vert \psi
_{+}^{x}\left( x\right) \right\vert ^{2}-\left\vert \psi _{-}^{x}\left(
x\right) \right\vert ^{2},  \label{SpinX-byWave}
$ and
$
\langle \sigma _{y}\left( x\right) \rangle  = i\left[ \psi _{-}^{z}\left(
x\right) ^{\ast }\psi _{+}^{z}\left( x\right) -\psi _{+}^{z}\left( x\right)
^{\ast }\psi _{-}^{z}\left( x\right) \right]  \label{SpinY-byWave}
$.
It turns out that the transitions have the nature of topological transitions
as the spin winding number~\cite{Ying-Spin-Winding,Ying-JC-winding,Ying-Topo-JC-nonHermitian} $n_{w}^{zx}$ in the $\langle \sigma _{z}\rangle $-%
$\langle \sigma _{x}\rangle $ plane has a transition at the EPs, as
illustrated by zero $n_{w}^{zx}$ in Fig.\ref{fig-topo-transition}(a) in the
PT/anti-PT-broken phases and finite $n_{w}^{zx}$ in
Fig.\ref{fig-topo-transition}(b) in the PT/anti-PT-symmetric phases. The spin winding is driven\cite{Ying-Spin-Winding,Ying-JC-winding,Ying-Topo-JC-nonHermitian} by the effective Rashba or Dresselhaus spin-orbit coupling similar\cite{Ying-2021-AQT,Ying-gapped-top,Ying-Stark-top,Ying-Spin-Winding,Ying-JC-winding,Ying-Topo-JC-nonHermitian,Larson-2010PRA-R} to nanowires\cite{Nagasawa2013Rings,Ying2016Ellipse,Ying2017EllipseSC,Ying2020PRR,Gentile2022NatElec} and
cold atoms\cite{Li2012PRL,LinRashbaBECExp2011}.
The zero $n_{w}^{zx}$ with vanishing $\langle \sigma
_{z}\left( x\right) \rangle $ comes from the equal amplitudes of $\psi _{\pm
}^{z}\left( x\right) $, due to purely real or imaginary components of
$\{C_{n\Uparrow }^{\left( \eta \right) },C_{n\Downarrow }^{\left( \eta
\right) }\}=\{\eta \sqrt{ng^{2}-\gamma ^{2}/4}-i\gamma /2,g\sqrt{n}\}$ and
$\{i(d_{\Omega \omega }/2+\eta \sqrt{d_{\Omega \omega }^{2}/4-n\Gamma ^{2}},
\Gamma \sqrt{n}\}$ ($d_{\Omega\omega}=\Omega-\omega$), respectively, in former phases, being different from
the complex $C_{n\Uparrow }^{\left( \eta \right) }$ in the latter phases.
The transition in $n_{w}^{zx}$ can be seen more clearly in a parameter variation
in Figs.\ref{fig-topo-transition}(c) and \ref{fig-topo-transition}(e) (solid blue lines),
and in 2D maps in Figs.\ref{fig-topo-transition}(c)
and \ref{fig-topo-transition}(e). The spin winding numbers $n_{w}^{yx}$
(dashed) $n_{w}^{zy}$ (dotted) in other planes in Figs.\ref%
{fig-topo-transition}(c) and \ref{fig-topo-transition}(e) have no change at
the EPs, although $n_{w}^{yx}$ reverses the sign (winding direction) at
resonance $\omega =\Omega $ away from the EPs.

{\it Simultaneous occurrence of LCT and TCT.}--Generally speaking, there are two
different classes of phase transitions, one is the LCT\cite{Landau1937} which breaks
symmetry, while the other class is TCT\cite{Topo-KT-transition,
Topo-KT-NoSymBreak,
Topo-Haldane-1,Topo-Haldane-2,
Topo-Wen,ColloqTopoWen2010,
Hasan2010-RMP-topo,Yu2010ScienceHall,Chen2019GapClosing,TopCriterion,Top-Song-2017,Top-Guan,TopNori,
Amaricci-2015-no-gap-closing,Xie-QAH-2021} which preserves or is protected by
symmetry. Conventionally they are incompatible due to the contrary
symmetry requirements. Here we see that the transitions at EPs are
simultaneously of both the symmetry-breaking LCT and the symmetry-protected
TCT. We summarize the characters of the transitions at the EPs in Table \ref{Table-LCT-TCT}.
As a key character of the LCT, the transitions at the
EPs break the symmetry of either PT or anti-PT. The symmetry expectation
amplitude $\left\vert \langle K\Pi _{x}\rangle \right\vert $ can be an
order parameter in the symmetry-broken phases which remains in a saturation
value $1$ in the symmetric phases but starts to decreases after the
transitions. The transitions are critical as indicated by the divergence of
the QFI which is equivalent to the fidelity susceptibility. On the other
hand, the transitions at the same time manifest the character of TCT in spin winding behaviors, which is protected by another parity
symmetry ${\cal P}=e^{i\pi a^{\dagger }a}\sigma _{x}$ \cite%
{Ying-2021-AQT,Ying-gapped-top,Ying-Stark-top,Ying-Spin-Winding,Ying-Topo-JC-nonHermitian,Ying-JC-winding}. As the key for transition-class reconciliation here, the simultaneous occurrence of the two conventionally
incompatible classes of transitions does not conflict in the symmetry requirements now,
as they involve different symmetries, $K\Pi _{x}$ and ${\cal P}$, whose
breaking and preserving are independent. Note that the spin winding number $%
n_{w}^{yx}$ is invariant in each phase, thus forming a topological
universality in contrast to the critical universality of the LCTs. Thus we
also hit two birds by one stone to have both classes of universalities at
the same time.

{\it Conclusions.}--We have analyzed the simultaneously
critical and topological features of transitions in the non-Hermitian JCMs,$\ H_{\gamma }$
and $H_{\Gamma }$, which have real energy spectra. $H_{\gamma }$ has the PT
symmetry while $H_{\Gamma }$ possesses the anti-PT symmetry in U(1)
subspace, the symmetry difference leads to opposite energy splitting and
degenerate behaviors in symmetric and symmetry-broken phases. However, the
QFI exhibits a critical character which is universal for different
parameters, all energy levels, both models, symmetric phases and symmetry-broken phases. The critical QFI may provide
sensitivity resource for the critical quantum metrology\cite{Garbe2020,Garbe2021-Metrology,Ilias2022-Metrology,Ying2022-Metrology,Montenegro2021-Metrology,Hotter2024-Metrology} and the universality
expands the metrology capability for various parameters and guarantees an
equally high measurement precision. Both the symmetry breaking aspect and
the critical and universal behavior characterize the LCT.
On the other hand, the transitions at the EPs also manifest character of TCT as the spin winding
number\cite{Ying-Spin-Winding,Ying-JC-winding,Ying-Topo-JC-nonHermitian} is zero in the PT-/anti-PT-broken phases while it is finite in the PT-/anti-PT-symmetric phases.
Thus the transitions at the EPs are
simultaneously LCTs\cite{Landau1937} and TCTs\cite{Topo-KT-transition,
Topo-KT-NoSymBreak,
Topo-Haldane-1,Topo-Haldane-2,
Topo-Wen,ColloqTopoWen2010,
Hasan2010-RMP-topo,Yu2010ScienceHall,Chen2019GapClosing,TopCriterion,Top-Song-2017,Top-Guan,TopNori,
Amaricci-2015-no-gap-closing,Xie-QAH-2021} which are conventionally
incompatible. Such a reconciliation of the two contradictory classes of
transitions stems from the fact that the TCTs are protected by the parity
symmetry which is different from the PT or anti-PT symmetry that the LCTs
break, which circumvents the contrary symmetry requirements of the LCTs and
the TCTs. Note that, unlike the first order transition in the
coexistence of the LCTs and the TCTs in Hermitian JCM\cite{Ying-JC-winding}, here the LCTs are
critical. Thus, we also have critical universality\cite{Sachdev-QPT,Irish2017,LiuM2017PRL,Ying-2021-AQT,Ying-Stark-top} and
topological universality\cite{Ying-2021-AQT,Ying-gapped-top,Ying-Stark-top,Ying-Spin-Winding,Ying-JC-winding,Ying-Topo-JC-nonHermitian} simultaneously. Besides establishing a paradigmatic case to break
the incompatibility of the LCTs and the TCTs in non-Hermitian systems, the
both availabilities of the sensitive critical feature and the robust
topological feature\cite{Ying-Topo-JC-nonHermitian} at the same time can also provide a particular potential
for designing more special quantum devices or sensors\cite{Garbe2020,Garbe2021-Metrology,Ilias2022-Metrology,Ying2022-Metrology,Montenegro2021-Metrology,Hotter2024-Metrology,
EP-sensors-Li-2023,EP-sensors-Miri-2019,EP-sensors-Ozdemir-2019,EP-sensors-Hokmabadi-2019,EP-sensors-Chen-2017,EP-sensors-Wiersig-2014,EP-sensors-Liu-2016} by simultaneously
making use of critical and topological advantages.

{\bf Acknowledgements}--This work was supported by the National Natural Science
Foundation of China through Grants No. 11974151 and No. 12247101.

\end{document}